\documentclass[10pt,a4paper,twoside]{article}

\usepackage{indentfirst}            
\usepackage{graphicx}               
\usepackage{amsgen,amsfonts,amssymb,amsbsy} 
\usepackage{amsmath}%

\newenvironment{resum}{\begin{quote}\small}{\end{quote}}

\newcommand{\bfsf}[1]{\textsf{\textbf{#1}}}

\begin{document}

\thispagestyle{plain}       

\begin{center}


{\LARGE\bfsf{MASS BOOM VERSUS BIG BANG:\\[15pt] EINSTEIN WAS RIGHT }}

\bigskip


\textbf{Antonio Alfonso-Faus }


\emph{E.U.I.T. Aeron\'autica \\
      Pl Cardenal Cisneros s/n 28040 Madrid, Spain\\
      \texttt{aalfonso@euita.upm.es}}\/

\end{center}

\medskip


\begin{resum}
When considering possible time variations of fundamental physical constants one has to keep
firm well established principles. Following this approach we keep firm the Action Principle,
General Relativity (the Equivalence Principle), and Mach's Principle. Also we introduce a new
principle under the name of ``TOTAL INTERACTION'' and reconsider Weinberg's relation with a
new approach. Consistent with all these principles we find that all masses increase linearly
with cosmological time (THE MASS BOOM) and that Planck's constant decreases also with this
time.Then the whole quantum world shrinks with time too. This is the cause of the red shift
(it is an alternative to the expansion of the Universe interpretation, and explains the BIG
BANG model approach as an apparent interpretation of the observers). The speed of light turns
out to be decreasing also with time. An ``absolute'' cosmological model arises, similar to the
one Einstein proposed, static, closed and finite, with the cosmological constant included. The
``relative'' model, the Universe as seen from the Lab observers, is an expanding one with a
quadratic law in time for the cosmological scale factor.

\end{resum}


\bigskip


\section{Introduction}

The beginning of scientific cosmology can be placed with the advent of the cosmological
equations of Einstein, as derived from his theory of general relativity. A Universe with
masses, and therefore gravitation, would be expected to contract, and that was the case
initially predicted. To avoid this, and have a static solution for the Universe as a whole,
Einstein included in his equations a cosmological constant, the lambda term, that resulted in
a push to balance gravitation. Then he obtained a static model, with curved space giving a
closed as well as finite Universe. Later Hubble found the redshift from distant galaxies to
increase with distance. One of the possible explanations was to consider that the Universe was
expanding as seen from any observer. Of course, going backwards in time, with this
interpretation the Universe would be seen as contracting, therefore would be initially
``born'' from a relatively small size and much hotter and denser than now. An expanding
Universe, found as a solution to the Einstein's equations, was proposed by Friedman and
therefore the scientific community had a theoretical frame to explain one interpretation of
Hubble's discovery. The concept of an expanding Universe from an initial Big Bang spread
rapidly and has been considered the best model, and therefore the best frame, to interpret all
current observations. Nevertheless this model has many problems and contains many paradoxes
that have been in part solved with additional theoretical inclusions (e.g. inflation).

In the present work we introduce the concept of a Mass Boom, already present in the
literature\cite{1} two years ago at the IV International Congress in Hyderabad, India, but now
we present here important modifications and refinements. Keeping first principles firm we
prove that all masses grow (increase) linearly with cosmological time. Conservation of
momentum implies then that the speed of light decreases linearly with time. It is seen that
Mach's principle, and its equivalent the principle of equivalence, give a unique solution for
the Universe that excludes expansion. We then reinterpret the redshift found by Hubble, and
prove that at the laboratory system it is proportional to Planck's ``constant'', a result that
comes from the comparison of frequencies. The new interpretation that we present here depends
exclusively on Planck's ``constant'', that we find decreasing with cosmological time. The
constancy of Planck's units of time and mass completely defines the time variation of the
Planck's constant. The new model we propose for the Universe is a static, closed and finite
one, as Einstein initially proposed. On the other hand we find a quantum world contracting
with cosmological time, in accordance with the decrease of Planck's constant, as interpreted
from the redshift. It is clear that if we take the quantum world as the reference, then the
Universe would be apparently expanding. It turns out that this apparent expansion is an
accelerated one, something already observed from the supernova type Ia measurements\cite{2}.
The apparent expansion we find is $a(t)\propto t^2$, for the cosmological scale factor $a(t)$.

We then solve the cosmological equations, and find the corresponding numerical values for the
dimensionless matter parameter $\Omega_m=1/3$ and the lambda parameter $\Omega_\lambda=1/3$,
which are very close to the values observed in many experiments. It turns out that the
apparent curvature term becomes rapidly negligible with age, so that practically we live in a
flat Universe, as seen from our Lab, also a well known current observation. The entropy of the
Universe is found to be similar to the cosmological time, as well as to the total matter of
the Universe. Then, we can talk of a Mass Boom as well as an Entropy Boom, equivalent to
cosmological time.

The natural units that emerge from this model, the ``true constants of nature'' are Planck's
mass and time, and of course the present size of the Universe $ct \approx 10^{28}$~cm, which
is the Planck's length at the first ``tic'' of the Universe (at the Planck's time).

Finally, the Pioneer 10/11 anomalous acceleration~\cite{3} observed is explained by our theory
presented here.

\section{The Mass Boom, Predicted by First Principles}

There have been doubts whether general relativity included Mach's principle or not. Certainly
it includes the equivalence principle, and now we will present an interpretation of both
principles that proves them to be equivalent. One interpretation of Mach's principle considers
the mass (energy) of a particle $m$ as due to its gravitational potential energy with respect
to the mass $M_u$ of the rest of the Universe
\begin{equation}\label{1}
  \dfrac{GM_um}{ct}\simeq mc^2
\end{equation}
General relativity is based upon the equivalence principle. One way to express it in
mathematical terms is to preserve, under any sort of time-variations, the ratio of the square
of any speed due to gravitation, $v^2 =GM/r$, to the square of the speed of light $c^2$, i.e.

\begin{equation}\label{2}
   \dfrac{v^2}{c^2}=\dfrac{GM}{c^2r}=\text{const.}
\end{equation}
The constancy of this ratio ensures the preservation of the principle of relativity under
cosmological time variations. If we substitute for the size $r$ the size of the seeable
Universe, $ct$, and for $M$ the mass of the Universe $M_u$, one gets
\begin{equation}\label{3}
  \dfrac{GM_u}{c^3t}=\text{const.}
\end{equation}
We see that the expressions (\ref{1}) and (\ref{3}) are equivalent. In the next section on the
action principle we prove that $G$ and $c^3$ have to be proportional to preserve the standard
form of the field equations of general relativity. The result is that the mass of the Universe
has to be proportional to the cosmological time (the Mass Boom):
\begin{equation}\label{4}
  M_u=\text{const}\cdot t
\end{equation}
We present now what we call the total interaction principle. It is a mathematical expression
that follows the requirement that all the gravitational interactions in the Universe must have
a mean free path, under a Newtonian point of view, of the order of the size of the Universe.
Then,
\begin{equation}\label{5}
  ct\simeq \dfrac{1}{n\sigma_g}
\end{equation}
where $n$ is the number density of particles in the Universe and $\sigma_g$ their
gravitational cross section as defined elsewhere\cite{4} and given by
\begin{equation}\label{6}
  \sigma_g=4\pi\dfrac{Gm}{c^2}\cdot ct
\end{equation}
Substituting the above into (\ref{5}) one has
\begin{equation}\label{7}
  ct\simeq \dfrac{(ct)^3}{\dfrac{G}{c2}M_u c t}
\end{equation}
i.e.
\begin{equation}\label{8}
  ct\simeq\dfrac{GM_u}{c^2}
\end{equation}
which is the same as (\ref{3}), the equivalence principle, and the same as (\ref{1}), the
Mach's principle.

Finally, by using the mass of the quantum of gravity $m_g$ defined elsewhere\cite{5}  as
\begin{equation}\label{9}
  m_g=\dfrac{\hbar}{c^2t}
\end{equation}
and calculating the mass rate of change dm/dt as given by the ratio $m_g/\tau$, where $\tau$
is the time for light to travel a Compton size $\hbar/mc$ one has:
\begin{equation}\label{10}
  \dfrac{dm}{dt}\simeq\dfrac{\hbar}{c^2t}\dfrac{mc^2}{\hbar}=\dfrac{m}{t}
\end{equation}
so that we get by integration
\begin{equation}\label{11}
   m=\text{const}\cdot t
\end{equation}
and therefore we obtain again the Mass Boom effect. Since $M_u$ and $m$ are proportional to
time, the number of particles of cosmological significance in the Universe is constant. The
time dependence corresponds to the mass. The above presentation has been submitted to
\emph{Physics Essays}\/ \cite{6}.

\section{The action principle}

Einstein's field equations can be derived from an action integral following the Least Action
Principle. In standard general relativity one has for the action integral\cite{7}:
\begin{equation}\label{12}
   \begin{array}{lcl}
     A & = & I_G+I_M\\
     A & = & -c^3/(16\pi G) \int R(g)^{1/2} d^4x + I_M
   \end{array}
\end{equation}
where $I_M$ is the matter action and $I_G$ the gravitational term. Then one obtains the field
equations
\begin{equation}\label{13}
  G^{\mu\nu}=8\pi(G/c^4)\cdot T^{\mu\nu}
\end{equation}
We assume a space-time metric and use the Robertson-Walker model that satisfies the Weyl
postulate and the cosmological principle, i.e.
\begin{equation}\label{14}
  ds^2  =   c(t)^2  dt^2  - R^2(t) \left\{ dr^2  / (1-kr^2)  + %
        r^2 \left( d\theta^2 + \sin^2\theta d\phi^2 \right) \right\}
\end{equation}
Einstein's equations (\ref{13}) follow from the Action (\ref{12}) provided that  the variation
of the coefficient in the integral  in equation (\ref{12})  be zero. Then
\begin{equation}\label{15}
  c^3/(16\pi G)=  \text{constant}
\end{equation}
We see that the assumption of a time varying $G$ must include a time varying $c$ to preserve
the form of the field equations.

The equation (\ref{15}) strongly suggests a specific link between mass and time. This is
\begin{equation}\label{16}
   c^3/G \simeq  4.04 \times 10^{38} \text{grams/sec} = \text{constant}
\end{equation}
which is of the order of the ratio of the mass of the observable Universe to its age.

On the other hand, the action for a free material point is
\begin{equation}\label{17}
  A= -mc \int ds
\end{equation}
To preserve standard mechanics we make the momentum $mc$ constant, independent from the
cosmological time, then
\begin{equation}\label{18}
  mc = \text{constant}
\end{equation}
With the constancies expressed in (\ref{15}) and (\ref{18}), general relativity is preserved
and of course the Newtonian mechanics too. Within these limits time variations of some of the
fundamental constants, $G$, $c$ and masses, are allowed at the same time preserving the laws
of physics as we know them today. From (\ref{18}) and the Mass Boom effect, the speed of light
\emph{decreases}\/ linearly with time $c \propto 1/t$ . It is evident that, with such a law
for the speed of light, the size of the Universe (of the order of $ct$) is constant and
therefore there is no ``absolute'' expansion.

\section{Reinterpretation of the Red Shift:  Time Variation of Planck's ``Constant''}

The ratio of frequencies observed at the laboratory system, photons from distant galaxies as
$\nu = c/\lambda$ and local atomic clocks as $v_0 \propto mc^2/\hbar$, with $mc$ constant and
$\lambda$ also constant (no expansion), gives a red shift proportional to $\hbar$. With no
expansion the red shift implies a \emph{decreasing}\/ Planck's ``constant''. Now, Planck's
units are defined as a combination of $G$, $c$ and $\hbar$:
\begin{equation}\label{19}
  \begin{array}{ll}
     \text{Planck's mass}     &    (\hbar c/G)^{1/2} = 2 \times 10^{-5} \text{ grams}\\ %
     \text{Planck's time}     &    (G\hbar/c^5)^{1/2} = 5.4 \times 10^{-44} \text{ sec}\\ %
     \text{Planck's length}   &    (G\hbar/c^3)^{1/2} = 1.6 \times 10^{-33} \text{ cm}\\ %
  \end{array}
\end{equation}
It is evident that if we choose a system of units such that $G = c^3$, as required by the Mass
Boom effect on the whole Universe, and such that $\hbar=c^2$, we get Planck's units of mass
and time as the ``natural'' units of mass and time. This is very appealing because the ratio
of the mass and age of the Universe to the corresponding Planck's units is the same factor of
about $10^{61}$. On the other hand the constant size of the Universe, the model we present
here, has a value of the order of $ct = 10^{28}$ cm, which is Planck's length at the first
``tic'' of time (at Planck's time).

\emph{\textbf{We see now that the Boom of an initial fluctuation of time and mass of the
Planck's units, by the same factor $10^{61}$, brings the fluctuation up to the state of the
Universe as we observe it today in time and mass. On the other hand the initial fluctuation
had a size of the order of Planck's length at that time, which is the constant size of the
Universe.} }\/

Then, this factor of $10^{61}$ is representative of the evolution of the initial fluctuation,
as characterized by the Planck's units, followed then by the Mass Boom to bring the Universe
to the present conditions. The magic number of the Universe is then $10^{61}$, as
representative of its evolution from the initial fluctuation up to now. The cosmology to be
studied now in this model is one that keeps $G = c^3$, $\hbar=c^2$ and $ct = 1$.

\section{Cosmological Equations}

The Einstein cosmological equations derived from his general theory of relativity are\cite{7}
\begin{equation}\label{20}
  \begin{array}{rcl}
  \left(\dfrac{\dot a}{a}\right)^2 + \dfrac{2\ddot a}{a} + 8\pi G\dfrac{p}{c2} + \dfrac{kc^2}{a^2} & %
     = & \Lambda c^2 \\[0.25cm] %
  \left(\dfrac{\dot a}{a}\right)^2 - \dfrac{8\pi}{3} G \rho + \dfrac{kc^2}{a^2} & %
     = & \dfrac{\Lambda c^2}{3}
  \end{array}
\end{equation}
The solution for $\hbar = c^2$, as presented in the previous section, implies a redshift given
by an apparent value of $a(t) \propto t^2$. In the units we have selected, consistent with
this interpretation of the redshift as a decrease in $\hbar$, the curvature term in (\ref{20})
decreases as $t^{-4}$ so that it is negligible, and we are observing essentially a flat
Universe. With the present reasonable approximation of zero pressure (neglecting random speeds
of galaxies), and substituting $a(t) \propto t^2$ in (\ref{20}) we finally get the
cosmological equations:
\begin{equation}\label{21}
  \begin{array}{rcl}
    \left(\dfrac{2}{t}\right)^2 + \dfrac{4}{t^2} & = & \Lambda c^2 \\[0.35cm] %
    \left(\dfrac{2}{t}\right)^2 - \dfrac{8\pi}{3}G\rho & = & \dfrac{\Lambda c^2}{3}
  \end{array}
\end{equation}
We convert now these equations to the standard definitions:
\begin{equation}\label{22}
  \begin{array}{ll}
  \Omega_m & = \dfrac{8\pi}{3} \dfrac{G\rho a^2}{\dot a^2} = %
      \dfrac{8\pi}{3} G \rho \dfrac{t^2}{4}\\[0.25cm] %
  \Omega_\Lambda & = \dfrac{\Lambda c^2}{3} \dfrac{a^2}{\dot a^2} = %
  \dfrac{\Lambda c^2}{3} \dfrac{t^2}{4}
  \end{array}
\end{equation}
and therefore we get
\begin{equation}\label{23}
  \begin{array}{lcl}
  \Omega_m & = & 1/3 \\
  \Omega_\Lambda & = & 2/3
  \end{array}
\end{equation}
These numbers are very close to the current values observed at present. The accelerated
expansion of the Universe~\cite{2} is then an apparent effect due to the quadratic relation
$a(t) \propto t^2$ as seen from the laboratory system.

\section{Entropy of the Universe: Linear with Time}

We have proved elswhere\cite{4} that the entropy of the Universe varies linearly with
cosmological time, based upon a new approach. However, using the well known
Bekenstein~\cite{9} and Hawking~\cite{10} relations for entropy, as well as the classical
definition, the result is the same: there is no escape, the entropy varies linearly with time
and for the Universe the high entropy of today is due to the fact that the Universe is very
old. \emph{\textbf{There is no entropy problem in our model.}}\/

Boltzmann constant $k$ varies in our theory as $c$. To see this we have the photon relation
typical for blackbody radiation
\begin{equation}\label{24}
  kT \propto \hbar c / \lambda
\end{equation}
Taking the laboratory system $\hbar$ is constant and from the empirical law $T\lambda
=$~constant  we get $k$ varying as $c$, inversely proportional to cosmological time. The
apparent time variation of $T$ is $T \propto 1/\lambda \propto 1/a(t)$. Hence the Bekenstein
definition of entropy:
\begin{align}\label{25}
  & S/k \propto \text{Energy} \times \text{size} / \hbar c  \propto  Mc^2 \times (ct) /\hbar c \propto t^2\\ %
  & \qquad\qquad \text{gives}\qquad S\propto t
\end{align}
For the Hawking black hole entropy: $S/k  \propto  1 /\hbar c \cdot (GM^2) \propto  1/\hbar  =
t^2$ i.e., the same result. For the standard    $S = Energy/T \propto Mc^2 \times a(t) \propto
t$ we also get the same result.

\section{The Magic Numbers}

The only magic number we found here is $10^{61}$ that brings the first Planck fluctuation to
the present state of the Universe. The Dirac magic number $10^{40}$, as the ratio of the size
of the Universe to the size of fundamental constants, and the ratio of electric to
gravitational forces, is a function of time in our approach here. Therefore the similarity of
these two values is a coincidence in our interpretation. Weinberg's  relation~\cite{7}, that
can be derived by equating the gravitational cross section (\ref{6}) of a particle of mass $m$
to the square of its Compton wavelength, is
\begin{equation}\label{26}
  \hbar^2/\left(Gct\right) \simeq m^3
\end{equation}
The time dependence implied here for a typical mass m of a particle is $m^3 \propto 1/t$ which
has no meaning in our approach. But at the Lab system we have $\hbar =\text{constant}$ and
then we get from (\ref{27}) that $m$ is proportional to time $t$, again the Mass Boom is also
present here.

\section{Predictions}

Using the expression of the fine structure constant found with no $c$ in it elsewhere~\cite{8}
we get
\begin{equation}\label{27}
  \alpha \simeq e^2/\hbar = \left(e/c\right)^2
\end{equation}
There have been no cosmologically significant time variations in $\alpha$, by that meaning
variations of the order of the variation of the cosmological age considered. Then one must
have $e/c = \text{constant}$, and therefore the electronic charge $e$ varies as $c$, inversely
proportional to $t$. However, in electromagnetic units $(e/c)$ is a true constant, so that the
Zeemann displacement is a constant in this theory, contrary to the statement made
elsewhere~\cite{1}. The apparent Hubble ``constant'' in this theory is $H = 2/t$, due to the
cosmological scale factor varying as $t^2$. Hence the Hubble age in this theory is twice as
much as the standard one. It is suggested that the age of the Universe may be as much as twice
what we have been thinking up to now. Finally the Pioneer~\cite{3} 10/11 anomalous
acceleration observed can be explained here by the ratio of the laboratory system reference
frequency ($\hbar = \text{constant}$) $\nu_1$
\begin{equation}\label{28}
  \nu_1 \propto mc^2/\hbar \propto c = 1/t
\end{equation}
and the frequency $\nu_p$ of the photon observed ($\hbar = c^2$)
\begin{equation}\label{29}
  \nu_p \propto mc^2/\hbar \propto m \propto t
\end{equation}

Hence we have $\nu_p / \nu_1 \propto t^2$. This is a \textbf{BLUE SHIFT}, as observed, and of
the order of $Hc \simeq 7 \times 10^{-8}$ cm/sec$^2$ to be compared with the observed value of
about $8 \times 10^{-8}$ cm/sec$^2$.

\section{Conclusions}

The Mass Boom proposed, linear increase of all masses with time, implies here a linear
decrease of the speed of light. The resultant cosmological model, static, almost flat, closed
and finite, has cosmological parameters in accordance with current observations. Main problems
of the standard model are solved: entropy, lambda constant, horizon etc. In fact many of these
problems are one and the same thing. Solving one you solve them all. This is the case here.

Finally, the time reversibility of all the equations of physics poses a deep theoretical
problem: nature has irreversible process, and this irreversibility is not now explicit in the
standard basic equations of physics (Newton's mechanics, quantum mechanics, general
relativity, etc). With our approach the Mass Boom ensures that irreversibility is present
everywhere: in fact we have proved that it corresponds to an Entropy Boom linear with time.


\end{document}